\documentstyle[12pt,floats,epsf,aps]{revtex}

\small
\begin{document}
\newcommand{\be}{\begin{equation}}
\newcommand{\ee}{\end{equation}}
\newcommand{\bea}{\begin{eqnarray}}
\newcommand{\eea}{\end{eqnarray}}

\title{
{\bf Kaon production at subthreshold and threshold energies } }
\author{J\"org Aichelin\footnote{invited speaker} and Christoph Hartnack\\
  SUBATECH \\
  Universit\'e de Nantes, EMN, IN2P3/CNRS \\
  4, Rue Alfred Kastler, 44070 Nantes Cedex 03, France} 
\maketitle
 
\begin{abstract}
We summarize what we have learnt about the kaon production in nucleus-nucleus
collisions in the last decade. We will address three questions:
a) Is the $K^+$ production sensitive to the nuclear equation of state? b)
How can it happen that at the same excess energy the same number of $K^+$ 
and $K^-$ are produced in heavy ion collisions although the elementary
cross section in pp collisions differs by orders of magnitudes? and 
c) Why kaons don't flow?
\end{abstract}

\section{Is the $K^+$ production sensitive to the nuclear equation of state?}
Already in the  eighties it has been observed in theoretical calculations 
\cite{aiko}
that the total $K^+$ production cross section at subthreshold or threshold
energies in collisions of heavy ions depends on the
nuclear equation of state (EOS). This has been confirmed in the mean time 
by many other groups. For light ions such an observation 
could not be made. Therefore it was suggested that the ratio of the 
production yield of $K^+$ in heavy and that in light systems can serve 
as a signal for the nuclear EOS.
 
Motivated by the fact that $K's$ are Goldstone bosons
it has been assumed in these calculations that their mass does not change in a
nuclear environment. Following earlier suggestions \cite{rako} it has been
further assumed that they are created in the baryonic reactions 
\be B+B\rightarrow \Lambda + K^+ N \ee where B is either a 
nucleon or a $\Delta$.
At the energies considered the pionic channel is negligible. 

Detailed studies have revealed that this dependence on the EOS can be easily
understood.  The softer the EOS the higher is the average density the 
nuclear system reaches in central collisions. As a consequence
the mean free path becomes shorter and therefore collisions are more
frequent. This effect is even amplified in the case of the $\Delta$: If the
density increases the chance that a $\Delta$ has a collision with 
another baryon before it disintegrates becomes higher.   

For the standard soft and hard EOS's \cite{aic,har} the $K^+$ yield differed
by a factor of two and hence it seemed possible to achieve an experimental
determination of the nuclear equation of state. However, the calculations 
were plagued by the poor knowledge of the elementary cross sections (eq.1) of
which at the relevant energies not even the order of magnitude 
was known experimentally. Theoretical calculation have just started to be
advanced at that time \cite{lag}.

In the mean time, thanks to an intensive program at COSY \cite{cosya,cosyb} the 
$pp \rightarrow \Lambda K^+ N$ and $pp \rightarrow \Sigma K^+ N$ cross sections
close sto the threshold have been measured. Although the relative ratio of 
both
is not yet understood, as far as $K^+$ mesons are concerned this is of no importance.
The problem is, however, how to extrapolate the cross section to np and nn
reactions. Depending on whether a kaon or a pion is exchanged between the
nucleons the isospin coefficients  differ up to a factor of 2.
The cross section including a $\Delta$ in the entrance channel have been
calculated theoretically \cite{tsu}. They await an experimental confirmation 
in pA
and $\pi$A experiments. Also the $\pi B \rightarrow \Lambda(\Sigma)K^+$
cross section is understood now theoretically \cite{tsua}. Using these cross
sections the simulation programs reproduce the experimental excitation
functions. However the error bars due to the uncertainties of the elementary
cross sections are still large. It is this incertitude which limits presently 
the predictive power of the simulation programs for heavy ion reactions, 
as we will see later.

Parallel to the experimental and theoretical investigation of the elementary
reaction cross sections detailed studies on the behavior of kaons in the medium
have been advanced \cite{scha}. Today very different approaches ( based on the
chiral perturbation theory\cite{was}, on the Nambu- Jona- Lasinio model
\cite{neb} and on a coupled
channel approach \cite{schaa} ) result almost in the same modification of the 
$K^+$ mass in
a nuclear environment as we can see in fig.1.
We  see also that for moderate densities the
mass of the $K$'s change about linearly with the density.
\begin{figure}[t]
\begin{tabular}{lcr} &
\begin{minipage}{14.2cm}
\epsfxsize=9.4cm
$$
\epsfbox{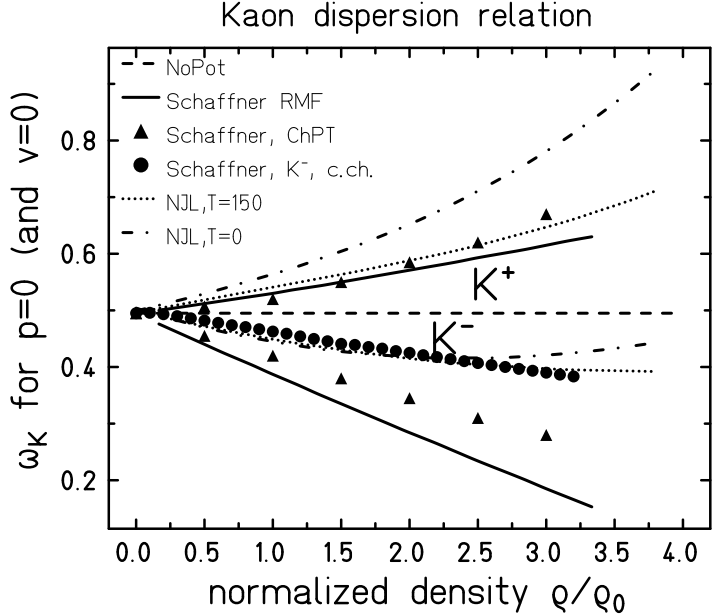}
$$
\caption{\textit{Energy of the kaons for zero momentum as a function of the
baryon density obtained in three different models, a mean field approach, chiral
perturbation theory(ChPT), a coupled channel calculation and the Nambu Jona Lasinio (NJL) model at two
different
temperatures of the matter}}
\label{wirk1}
\end{minipage} & 
\end{tabular}
\end{figure}

The mass of the $K^+$ increases by about 35 MeV per density unit whereas that of
the $K^- $ decreases with about the same amount. 

Despite of its smallness the change of the mass of the $K^+$'s in a nuclear
environment has rendered the kaons useless for a experimental determination of
the EOS \cite{hir}. The larger collision frequency for a soft EOS is counterbalanced by a
larger threshold due to a larger kaon mass and consequently by a 
smaller production cross section. The small
differences which are still observed in the simulation programs between a soft
and a hard EOS are too small in order to surmount the uncertainties imposed by
the limited knowledge on the elementary cross section. 

How the change of the kaon mass in medium and the incertitude of the elementary
cross sections influence the kaon yield obtained in the simulation programs for
heavy ion collisions is shown in fig.2 and 3, see as well  \cite{cro}. 
  
In fig. 2 we compare  the rapidity distributions measured by the FOPI 
\cite{rit} and the KAOS \cite{oes} collaborations 
with several IQMD calculations \cite{har} using a different description of
the kaons in the medium. In the No Pot calculations it is assumed that the
properties of the kaons are not modified in the medium.  The other three
calculations refer to the change of the kaon mass as displayed in fig. 2.  
In fig. 3 we have exchanged our cross section set ($\sigma$ Nan) 
for the kaon producing 
channels by the that of the Giessen group ($\sigma$ GI)\cite{gie} 
in the otherwise unchanged time evolution of the
IQMD approach. We see a change of about 30\% of the dN/dy at midrapidity, which
is solely caused by the different cross section parametrisations. This can be
verified by a comparison with the results published by the Giessen group
(fig. 5.5 of ref.\cite{gie}). For the same kaon production cross section both
calculations agree quantitatively. This shows that the overall dynamics of the
heavy ion reactions is almost identical in these both completely independent 
simulation programs. One may therefore conjecture that it is well under control. 
\begin{figure}[hbt]
\begin{tabular}{lcr}
\begin{minipage}{8.2cm}
\epsfxsize=8.cm
$$
\epsfbox{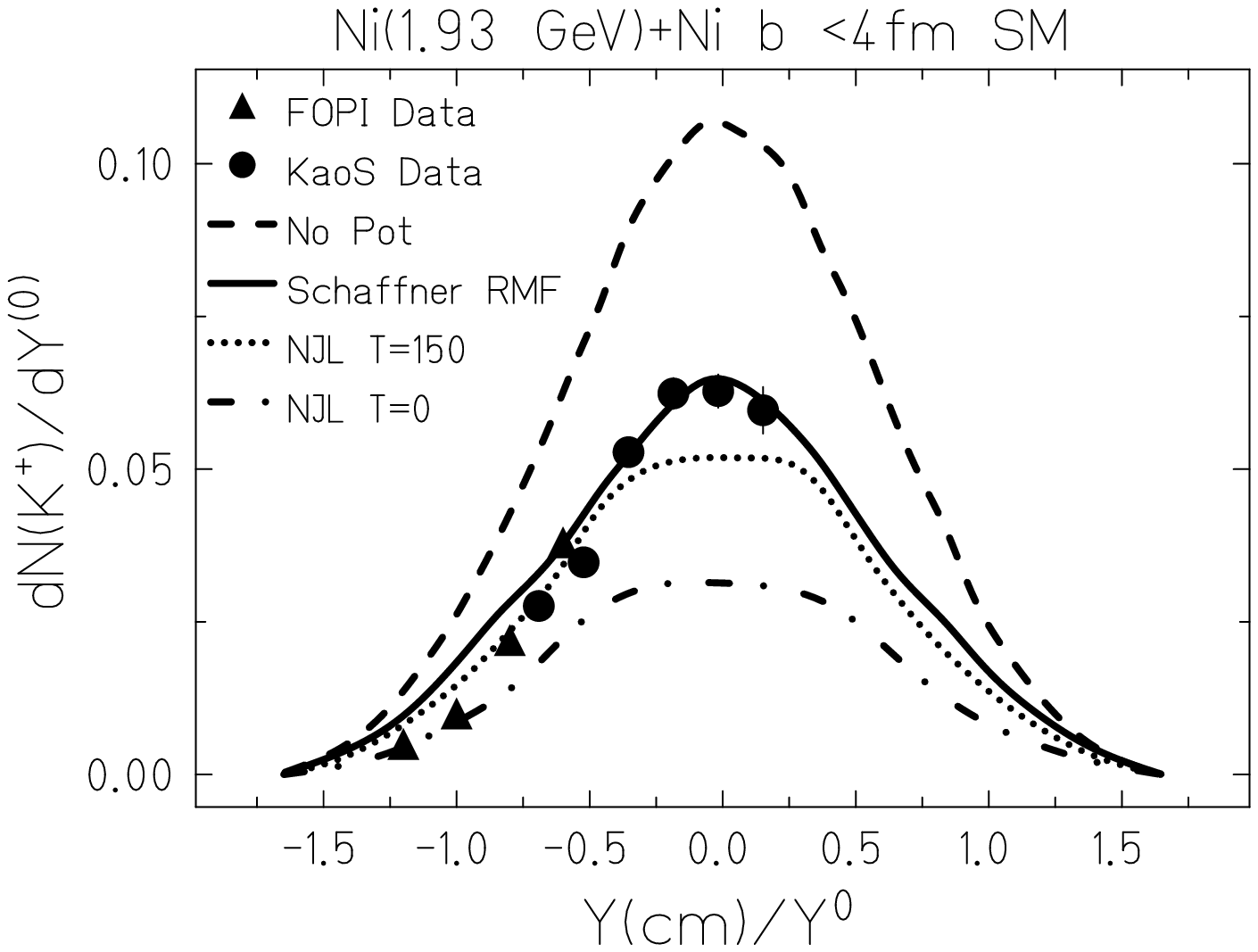}
$$
\caption{\textit{ The rapidity distribution of 
$K^+$'s for Ni(1.93 AGeV) + Ni, central collisions, measured by the FOPI
and KaoS collaborations as compared with the 
calculations for different kaon potentials }}
\label{wirk2}
\end{minipage} & &
\begin{minipage}{8.2cm}
\epsfxsize=8.cm
$$
\epsfbox{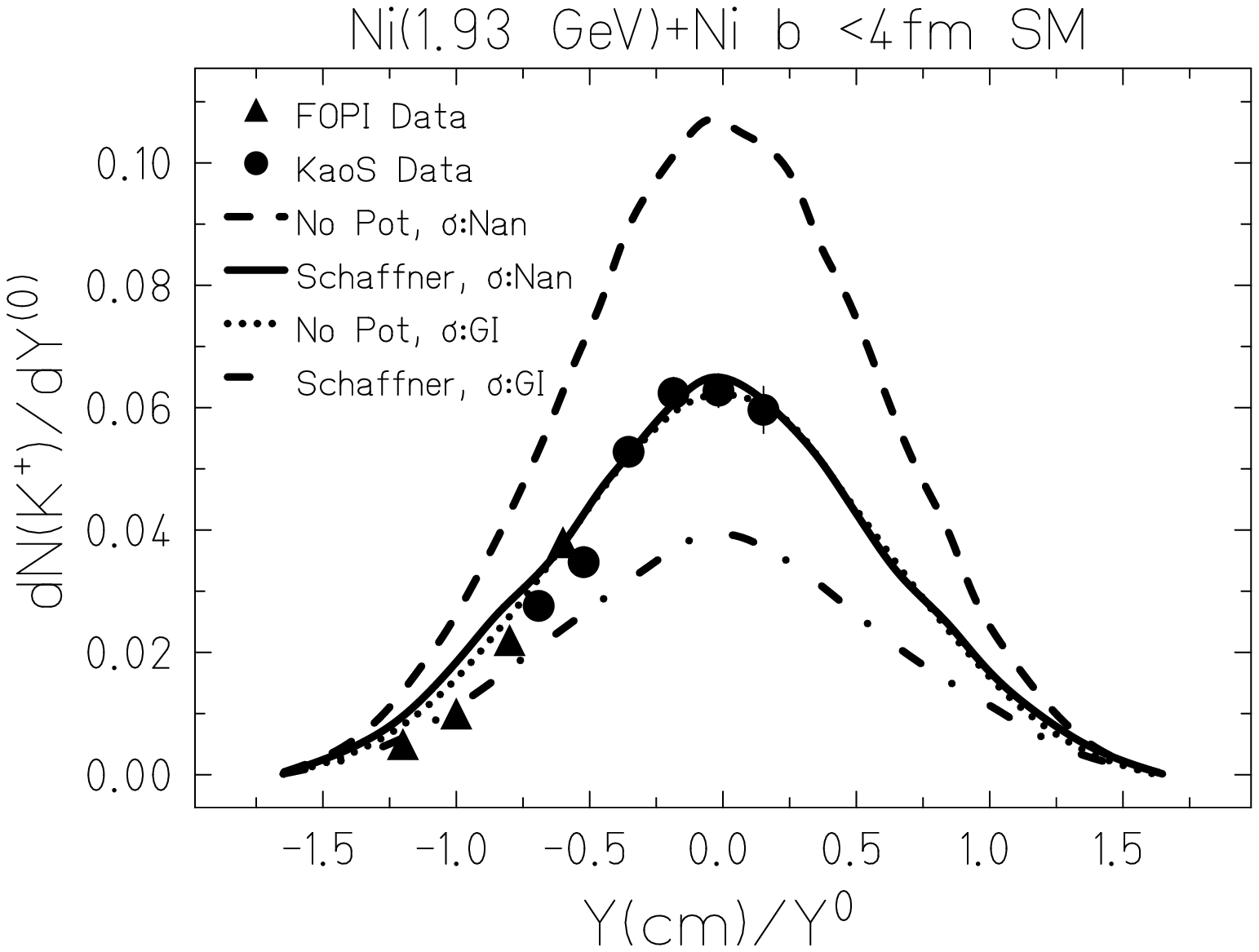}
$$
\caption{\textit{ The rapidity distribution of 
$K^+$'s for Ni(1.93 AGeV) + Ni, central collisions, measured by the FOPI
and KaoS collaborations as compared 
with the calculations using different production
cross sections}}
\end{minipage}
\end{tabular}
\end{figure}

\section{How can it happen that at the same excess energy the same number of
$K^+$ and $K^-$ are produced although the elementary
cross section in pp collisions differs by orders of magnitudes?}

Three years ago the KAOS collaboration has confronted us with the fact that in
Ni + Ni reactions $K^-$ and  $K^+$ meson are produced with the same probability
if the difference between the beam energy per nucleon and the production 
threshold is the same. This is very surprising because the elementary cross sections
$pp \rightarrow K^+\Lambda(\Sigma)N$ and $pp \rightarrow K^-K^+pp$ differ by several orders of
magnitude close to the threshold as a function of $\sqrt{s} - \sqrt{s_{thres}}$.
This puzzle has been resolved only recently. $K^-$ mesons produced at high
density in initial $NN \rightarrow NNK^+K^-$ reactions
have almost no chance to escape from the reaction zone without being
absorbed. This is due to the strongly exothermic reaction $K^-N 
\rightarrow \Lambda (\Sigma)\pi$ which has an appreciable cross section.
Hence the $K^-$ observed in the detector do not come from the initial collisions
but are produced in the inverse reaction $\Lambda (\Sigma) \pi\rightarrow K^- N$
at low densities, where the mass change is negligible.
Indeed, during the expansion an equilibrium is built up in this reaction
channel. This secondary interactions are absent in the $K^+$ channel.
Consequently, the variable $\sqrt{s} - \sqrt{s_{thres}}$ which make reference to
the $NNK^+K^-$ channel is not relevant at all and the agreement of the kaon
yields at the same excess energies has to be considered as accidental.

Fig. 4 displays the influence of the in medium properties of the kaons on the
observed $K^-$ yield where the spectra obtained for the systems 
C+C, Ni+Ni and Au+Au at different energies are compared to data from the
KaoS collaboration \cite{lau}. 
We observe the largest cross section if the kaons have
their free mass (NoPot). This is astonishing. One could believe that the
reduction of the $K^-$ mass in the medium increases the $K^-$ yield because the
threshold is lower. This is true of course but this effect is counterbalanced by
a strong decrease the $\Lambda's (\Sigma's)$ because less $K^+$'s are
produced if the mass of the $K^+$ increases in the medium. Independent of how 
the kaon mass is modified in the medium (see fig. 2) the final $K^-$ yield 
is identical in between the error bars as long as the $K^+$ masses are not
changed.  This is due to the fact that the finally observed $K^-$'s are 
created at low density where the mass differences are small.
\begin{figure}[t]
\begin{tabular}{c}
\begin{minipage}{16.6cm}
\epsfxsize=16.3cm
$$
\epsfbox{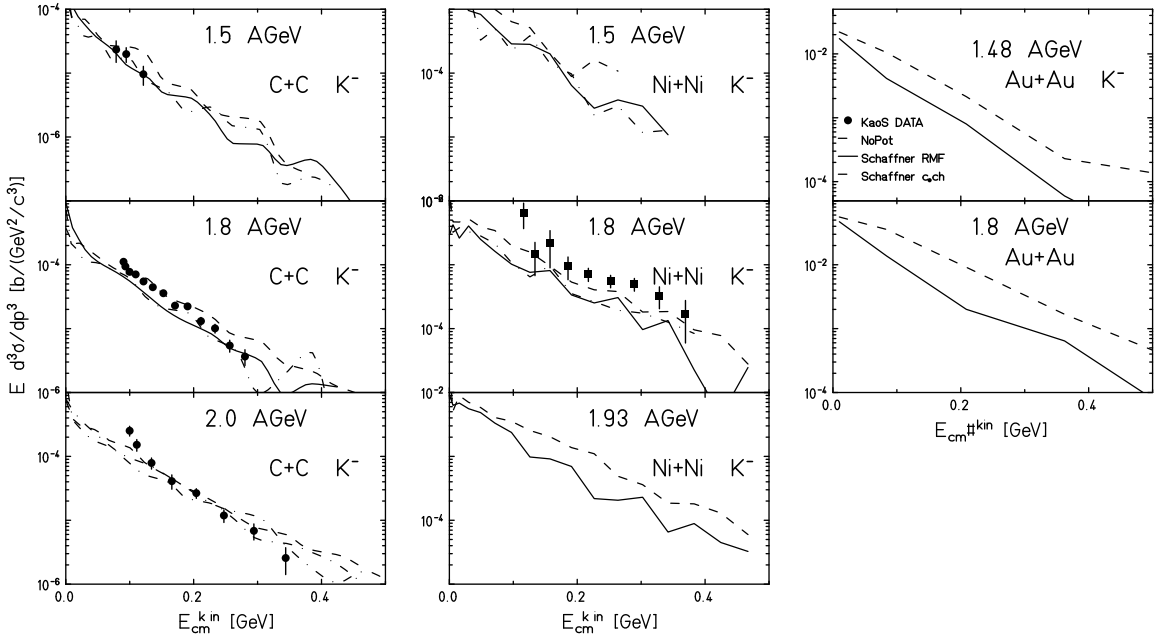}
$$
\caption{\textit{Double differential cross section of $K^-$ meson in C+C, 
Ni+Ni  and Au+Au reactions at different energies. The experimental data 
are confronted with
IQMD calculations employing different $K^-$ dispersion relations. }}
\label{km}
\end{minipage}
\end{tabular}
\end{figure}

\section{Why don't kaons flow?}
The in plane flow of kaons \cite{rit} has gained in the past a lot of interest
because it is very small ( as compared to that of the protons) and it 
was claimed that its observation allows a distinction between a 
vector potential and a scalar potential of the kaons\cite{ko}. It is, 
however, not evident what one can learn from separating vector and scalar potential
because both are large and have opposite sign and only both together 
describe the behavior of kaons in the medium.
The interesting quantity to look at is the
comparison between the in-plane flow with and without interaction of the $K^+$
with the nuclear medium. This effect is quite small. 

Much more interesting is the question why the in-plane flow $K^+$'s is that small,
a phenomenon which has been also addressed by the AGS collaborations. Being
produced in the elementary pp collisions the in-plane velocity of the $K^+$ 
should equal that of the protons or $\Lambda's$. This is obviously not the case. There
are two reasons \cite{dav} : 1) Nucleons which pass the high density zone 
(where the $K^+$'s are dominantly produced) have only half of the in-plane flow 
observed by averaging over all nucleons. This explains why the $\Lambda$
in-plane flow is only half of that of the protons. 2) In addition, the three body
phase space distribution places the $K^+$'s at rapidities which are far from the
rapidity of the source which determines the in-plane flow. Hence at a given 
rapidity kaons come from very different source rapidities and hence have 
very different in-plane flows. The negative and positive contributions cancel
almost and renders the net flow small. 

\section{Perspectives for AGS and SPS}
We have discussed three major results obtained in the last years in the field of
kaon production close to threshold energies. Which relevance have these results
at higher energies, at the center of interest of this conference? I see three
perspectives for heavy ion collisions at higher energies:
A) The search for
the kaon flow and the hope to learn something about the properties of the
surrounding matter is, as we have learnt, a hot topic as well at AGS energies. 
There the theoretical challenge is much larger in view of the many mesons and 
resonances produced in the reaction. Also the puzzle that different particles
show a different in-plane flow continues into this energy domain and the
presented results at lower energies may help for an understanding. 
B) We have seen that the kaon production close to threshold can only be described
by simulation programs if one assumes that the masses of the kaons change in the
medium. Otherwise we miss the experimental cross sections by about a factor of
2. The simulation programs for SPS and AGS energies presented at this conference
like URQMD or NEXUS are able to reproduce quantitatively the observed spectra 
without invoking mass changes. This is very puzzling because the density
obtained at AGS or SPS is still higher than that seen at SIS.
C) We have learnt that from these threshold energies up to the highest energies
available today the yield of particles, and especially that of the kaons, 
can be well described in thermal or statistical models\cite{oes}.  This is very puzzling 
because at these low energies the rapidity distribution of kaons is incompatible
with the assumption of thermal equilibrium. If this is not only an accident the
reactions at threshold energies which are much easier to access theoretically
may be useful to understand the physics behind this observation.

\end{document}